Low-lying exotic mesons in the coupled-channel formalism.

Gerasyuta S.M. [1,2], Kochkin V.I. [1]

1. Department of Theoretical Physics, St. Petersburg State University,
   198904, St. Petersburg, Russia.
2. Department of Physics, LTA, 194021, St. Petersburg, Russia.
   E-mail:gerasyuta@SG6488.spb.edu

Abstract.

The relativistic four-quark equations are found in the framework of the dispersion relation technique. The dynamical mixing of the four-quark amplitudes and the glueball amplitudes is considered. The approximate solutions of these equations using the method based on the extraction of leading singularities of the amplitudes are obtained. The four-quark amplitudes of exotic mesons including the quarks of three flavors ($u$, $d$, $s$) are calculated. The poles of these amplitudes determine the masses of the exotic mesons.

I. Introduction.

Developing an understanding of the forces between mesons is important in the analyzing of many anomalous effects in the low-energy meson physics. The first extensive calculation was by Jaffe [1] who applied a semiclassical approximation of the MIT-bag model to $qq\bar{q}\bar{q}$ spectroscopy. Jaffe and Low introduced a quantity, they called the P-matrix, whose poles are expected to correspond to the energies calculated in the bag model [2].

Weinstein and Isgur searched for the full four-quark wavefunction in a large variational Gaussian basis [3]. They inverted the Shrodinger equation to obtain an effective potential and integrated this to calculate the phase shift. The flux tube model [4], which is based on strong coupling Hamiltonian lattice gauge theory, has clarified the status of the $\vec{F}_i \cdot \vec{F}_j$ model [5]. The flux tube model demonstrates that the two dimensional color basis of the $\vec{F}_i \cdot \vec{F}_j$ model is a truncation of the complete flux tube basis. The truncation is a good one in the sence that it closely reproduces the two lowest adiabatic potentials of the flux tube model [6]. It should be stressed that all predictions of multiquark bound states are entirely model-dependent, thus the predictions of the quark potential model serve as an important test [7-10]. In series of papers [11-15], a practical treatment of relativistic three-hadron systems has been developed. The physics of the three-hadron system is usefully described in terms of the pairwise interactions among the three particles. The theory is based on the two principles of unitarity and analyticity, as applied to the two-body subenergy channels. The linear integral equations in a single variable are obtained for the isobar amplitudes. Instead of the quadrature, methods of obtaining the set of suitable functions are identified and used as a basis for the expansion of the desired solutions. By this means the coupled integral equations are solved in terms of simple algebra. In the recent papers [16-18], the relativistic three-quark equations for the excited baryons are found in the framework of the dispersion relations technique. We have used the orbital-spin-flavor wave functions for the construction of integral equations. We searched for the approximate solution of integral three-quark equations by taking into account two-particle and triangle singularities, all the weaker ones being neglected. If we considered such an approximation, which corresponds to taking into account two-body and triangle singularities, and defined all the smooth functions in the middle point the physical region of the Dalitz plot, then the problem was reduced to solving a system of simple algebraic equations.

We calculated the mass spectra of excited baryons using the input four-fermion interaction with the quantum numbers of gluon [5]. In the present paper the relativistic four-quark equations are found in the framework of coupled-channel formalism. The dynamical mixing between the glueball and the four-quark states is considered. The four-quark amplitudes of the exotic mesons are calculated.

In Section II the relativistic four-quark equations are constructed in the form of the dispersion relation over the two-body subenergy. The approximate solutions of these equations using the method based on the extraction of leading singularities of the amplitude are obtained. The four-quark amplitudes of low-lying exotic mesons are calculated. Section III is devoted to the calculation results for the exotic meson mass spectra (Tables I, II). In the Conclusion the status of the considered model is discussed.

II. Four-quark amplitudes.

We derive the relativistic four-quark equations in the framework of the dispersion relation technique. We use only planar diagrams; the other diagrams due to the rules of $1/N_c$ expansion [19-21] are neglected. The current generates a four-quark system. The correct equations for the amplitude are obtained by taking into account all possible subamplitudes. It corresponds to the division of complete system into subsystems with smaller number of particles. Then one should represent a four-particle amplitude as a sum of six subamplitudes:

$$A = A_{12} + A_{13} + A_{14} + A_{23} + A_{24} + A_{34}. \qquad (1)$$

This defines the division of the diagrams into groups according to the certain pair interaction of particles. The total amplitude can be represented graphically as a sum of diagrams. We need to consider only one group of diagrams and the amplitude corresponding to them, for example $A_{12}$. We shall consider the derivation of the relativistic generalization of the Faddeev-Yakubovsky approach. The set of diagrams associated with the amplitude $A_{12}$ can further be broken down into groups corresponding to subamplitudes: four-quark amplitudes $A_l(s,s_{12},s_{123})$ ($l$=1-4) and glueball amplitude $A_5(s,s_{12},s_{34})$ (Fig. 1). Here $s_{ik}$ is the two-particle subenergy squared, $s_{ijk}$ corresponds to the energy squared of particles $i$, $j$, $k$ and $s$ is the system total energy squared.

The system of graphical equations is determined by the subamplitudes using the self-consistent method. The coefficients are determined by the permutation of quarks [22, 23].

In order to represent the subamplitudes $A_l(s,s_{12},s_{123})$ ($l$=1-4) and $A_5(s,s_{12},s_{34})$ in the form of a dispersion relation it is necessary to define the amplitudes of quark-antiquark interaction $a_n(s_{ik})$. The pair quarks amplitudes $q\bar{q} \to q\bar{q}$ are calculated in the framework of the dispersion N/D method with the input four-fermion interaction [24-26] with the quantum numbers of the gluon [5, 27]. The regularization of the dispersion integral for the D-function is carried out with the cutoff parameter $\Lambda$.

The four-quark interaction is considered as an input:

$$g_v(\bar{q}\lambda I_f \gamma_\mu q)^2 + 2g_v^{(s)}(\bar{q}\lambda \gamma_\mu I_f q)(\bar{s}\lambda \gamma_\mu s) + g_v^{(ss)}(\bar{s}\lambda \gamma_\mu s)^2 \qquad (2)$$

Here $I_f$ is the unity matrix in the flavor space (u, d), $\lambda$ are the color Gell-Mann matrices. Dimensional constants of the four-fermion interaction $g_v$, $g_v^{(s)}$ and $g_v^{(ss)}$ are parameters of the model. At $g_v = g_v^{(s)} = g_v^{(ss)}$ the flavor $SU(3)_f$ symmetry occurs. The strange quark violates the flavor $SU(3)_f$ symmetry. In order to avoid additional violation parameters we introduce the scale shift of the dimensional parameters [27]:

$$g = \frac{m^2}{\pi^2} g_v = \frac{(m+m_s)^2}{4\pi^2} g_v^{(s)} = \frac{m_s^2}{\pi^2} g_v^{(ss)}, \qquad (3)$$

$$\Lambda = \frac{4\Lambda(ik)}{(m_i + m_k)^2}. \qquad (4)$$

Here $m_i$ and $m_k$ are the quark masses in the intermediate state of the quark loop. Dimensionless parameters g and $\Lambda$ are supposed to be constants which are independent of the quark interaction type. The applicability of Eq. (2) is verified by the success of De Rujula-Georgi-Glashow quark model [5], where only the short-range part of Breit potential connected with the gluon exchange is responsible for the mass splitting in hadron multiplets.

We use the results of our relativistic quark model [27] and write down the pair quark amplitudes in the form:

$$a_n(s_{ik}) = \frac{G_n^2(s_{ik})}{1-B_n(s_{ik})}, \qquad (5)$$

$$B_n(s_{ik}) = \int_{(m_i+m_k)^2}^{(m_i+m_k)^2 \Lambda/4} \frac{ds'_{ik}}{\pi} \frac{\rho_n(s'_{ik})G_n^2(s'_{ik})}{s'_{ik}-s_{ik}}. \qquad (6)$$

Here $G_n(s_{ik})$ are the quark-antiquark vertex functions (Table III). The vertex functions are determined by the contribution of the crossing channels. The vertex functions satisfy the Fierz relations. All of these vertex functions are generated from $g_v$, $g_v^{(s)}$ and $g_v^{(ss)}$. $B_n(s_{ik})$ and $\rho_n(s_{ik})$ are the Chew-Mandelstam functions with cutoff $\Lambda$ [28] and the phase space, respectively:

$$\rho_n(s_{ik}, J^{PC}) = \left(\alpha(n, J^{PC})\frac{s_{ik}}{(m_i+m_k)^2} + \beta(n, J^{PC}) + \delta(n, J^{PC})\frac{(m_i-m_k)^2}{s_{ik}}\right) \times$$
$$\times \frac{\sqrt{[s_{ik}-(m_i+m_k)^2][s_{ik}-(m_i-m_k)^2]}}{s_{ik}}, \qquad (7)$$

The coefficients $\alpha(n)$, $\beta(n)$ and $\delta(n)$ are given in Table IV.

Here n=1 corresponds to a $q\bar{q}$-pairs with $J^{PC}=1^{--}$ in the $8_c$ color state, n=2 defines the $q\bar{q}$-pairs corresponding to mesons with quantum numbers: $J^{PC}=0^{++},0^{-+},0^{--},1^{++},1^{-+},1^{--},2^{++}$.

In the case in question the interacting quarks do not produce a bound state, therefore the integration in Eqs. (8) - (12) is carried out from the threshold $(m_i+m_k)^2$ to the cutoff $\Lambda(ik)$. The coupled integral equation systems, corresponding to Fig. 1 (the meson state with n=2 and $J^{PC}=2^{++}$) can be described as:

$$A_1(s, s_{12}, s_{123}) = \frac{\lambda_1 B_2^{uu}(s_{12})}{1-B_2^{uu}(s_{12})} + 2\frac{G_1(s_{12})}{1-B_2^{uu}(s_{12})}[\hat{J}_1 A_1(s, s'_{13}, s_{123}) + $$
$$+ \hat{J}_3 A_5(s, s'_{13}, s'_{24})] \qquad (8)$$

$$A_2(s, s_{12}, s_{123}) = \frac{\lambda_2 B_2^{ss}(s_{12})}{1 - B_2^{ss}(s_{12})} + 2\frac{G_2(s_{12})}{1 - B_2^{ss}(s_{12})}[\hat{J}_1^{ss} A_2(s, s_{13}', s_{123}) + \quad (9)$$
$$+ \hat{J}_3^{ss} A_5(s, s_{13}', s_{24}')]$$

$$A_3(s, s_{12}, s_{123}) = \frac{\lambda_3 B_2^{us}(s_{12})}{1 - B_2^{us}(s_{12})} + 2\frac{G_2(s_{12})}{1 - B_2^{us}(s_{12})} \hat{J}_3^{s} A_5(s, s_{13}', s_{24}'), \quad (10)$$

$$A_4(s, s_{12}, s_{123}) = \frac{\lambda_4 B_2^{ud}(s_{12})}{1 - B_2^{ud}(s_{12})} + 2\frac{G_2(s_{12})}{1 - B_2^{ud}(s_{12})} \hat{J}_3 A_5(s, s_{13}', s_{24}'), \quad (11)$$

$$A_5(s, s_{12}, s_{34}) = \frac{\lambda_5 B_1(s_{12}) B_1(s_{34})}{[1 - B_1(s_{12})][1 - B_1(s_{34})]} +$$
$$+ 4\frac{G_1(s_{12}) G_1(s_{34})}{[1 - B_1(s_{12})][1 - B_1(s_{34})]}[\hat{J}_2 A_1(s, s_{13}', s_{134}') + \hat{J}_2 A_4(s, s_{13}', s_{134}') +, \quad (12)$$
$$+ \hat{J}_2^{s} A_3(s, s_{13}', s_{134}') + \hat{J}_2^{ss} A_2(s, s_{13}', s_{134}')]$$

$$B_1(s_{ik}) = \frac{1}{3}[B_1(s_{ik}, m_u, m_u) + B_1(s_{ik}, m_d, m_d) + B_1(s_{ik}, m_s, m_s)], \ i, k = 1, 2, 3, 4.$$

where $\lambda_i$ are the current constants. We introduce the integral operators:

$$\hat{J}_i = \hat{J}_i(s; m; m; m; m)$$
$$\hat{J}_i^{ss} = \hat{J}_i(s; m_s; m_s; m_s; m_s), \ i=1, 2, 3$$
$$\hat{J}_2^{s} = \hat{J}_2(s; m; m; m_s; m_s), \ \hat{J}_3^{s} = \hat{J}_3(s; m; m_s; m; m_s),$$

$$\hat{J}_1(s; m_1; m_2; m_3; m_4) = \int\limits_{(m_1+m_2)^2}^{(m_1+m_2)^2 \Lambda/4} \frac{ds_{12}'}{\pi} \frac{\rho_2(s_{12}') \cdot G_2(s_{12}')}{s_{12}' - s_{12}} \int\limits_{-1}^{+1} \frac{dz_1}{2}, \quad (13)$$

$$\hat{J}_2(s; m_1; m_2; m_3; m_4) = \int\limits_{(m_1+m_2)^2}^{(m_1+m_2)^2 \Lambda/4} \frac{ds_{12}'}{\pi} \frac{\rho_1(s_{12}') \cdot G_1(s_{12}')}{s_{12}' - s_{12}} \int\limits_{(m_3+m_4)^2}^{(m_3+m_4)^2 \Lambda/4} \frac{ds_{34}'}{\pi} \times$$
$$\times \frac{\rho_1(s_{34}') \cdot G_1(s_{34}')}{s_{34}' - s_{34}} \int\limits_{-1}^{+1} \frac{dz_3}{2} \int\limits_{-1}^{+1} \frac{dz_4}{2} \quad (14)$$

$$\hat{J}_3(s;m_1;m_2;m_3;m_4) = \frac{1}{4\pi} \times$$

$$\times \int_{(m_1+m_2)^2}^{(m_1+m_2)^2 \Lambda/4} \frac{ds'_{12}}{\pi} \frac{\rho_2(s'_{12}) \cdot G_2(s'_{12})}{s'_{12} - s_{12}} \int_{-1}^{+1} \frac{dz_1}{2} \int_{-1}^{+1} dz \int_{z_2^-}^{z_2^+} dz_2 \frac{1}{\sqrt{1 - z^2 - z_1^2 - z_2^2 + 2zz_1z_2}}, \quad (15)$$

where $m_i$ is a quark mass.

In Eqs. (13) and (15) $z_1$ is the cosine of the angle between the relative momentum of the particles 1 and 2 in the intermediate state and the momentum of the particle 3 in the final state, taken in the c.m. of particles 1 and 2. In Eq. (15) $z$ is the cosine of the angle between the momenta of the particles 3 and 4 in the final state, taken in the c.m. of particles 1 and 2. $z_2$ is the cosine of the angle between the relative momentum of particles 1 and 2 in the intermediate state and the momentum of the particle 4 in the final state, is taken in the c.m. of particles 1 and 2. In Eq. (14): $z_3$ is the cosine of the angle between relative momentum of particles 1 and 2 in the intermediate state and the relative momentum of particles 3 and 4 in the intermediate state, taken in the c.m. of particles 1 and 2. $z_4$ is the cosine of the angle between the relative momentum of the particles 3 and 4 in the intermediate state and that of the momentum of the particle 1 in the intermediate state, taken in the c.m. of particles 3, 4.

We can pass from the integration over the cosines of the angles to the integration over the subenergies.

Let us extract two-particle singularities in the amplitudes $A_l(s, s_{12}, s_{123})$ ($l$ =1-4) and $A_5(s, s_{12}, s_{34})$:

$$A_l(s, s_{12}, s_{123}) = \frac{\alpha_l(s, s_{12}, s_{123}) B_2(s_{12})}{1 - B_2(s_{12})}, \quad (16)$$

$$A_5(s, s_{12}, s_{34}) = \frac{\alpha_5(s, s_{12}, s_{34}) B_1(s_{12}) B_1(s_{34})}{[1 - B_1(s_{12})][1 - B_1(s_{34})]}, \quad (17)$$

We do not extract three- particle singularities, because they are weaker than two-particle singularities.

We used the classification of singularities, which was proposed in paper [29]. The construction of approximate solution of Eqs. (8) - (12) is based on the extraction of the leading singularities of the amplitudes. The main singularities in $s_{ik} \approx (m_i + m_k)^2$ are from pair rescattering of the particles i and k. First of all there are threshold square-root singularities. Also possible are pole singularities which correspond to the bound states. The diagrams of Fig.1 apart from two-particle singularities have the triangular singularities and the singularities defining the interaction of four particles. Such classification allows us to search the corresponding solution of Eqs. (8) - (12) by taking into account some definite number of leading singularities and neglecting all the weaker ones. We consider the approximation which defines two-particle, triangle and four-particle singularities. The functions $\alpha_l(s, s_{12}, s_{123})$ ($l = 1$-4) and $\alpha_5(s, s_{12}, s_{34})$ are the smooth functions of $s_{ik}$, $s_{ikl}$, $s$ as compared with the singular part of the amplitudes, hence they can be expanded in a series in the singularity point and only the first term of this series should be employed further. Using this classification, one defines the reduced amplitudes $\alpha_1$, $\alpha_2$, $\alpha_3$, $\alpha_4$, $\alpha_5$ as well as the B-functions in the middle point of the physical region of Dalitz-plot at the point $s^0$:

$$s_{1i}^0 = s_{j4}^0 = \frac{s + 8m^2}{6}, \quad s^0 = \frac{4[s - 2s_{1i}^0 + 2(m_1^2 + m_2^2 + m_3^2 + m_4^2)]}{(m_1 + m_j)^2 + (m_1 + m_4)^2 + (m_2 + m_3)^2 + (m_i + m_4)^2},$$

$$s_{123} = s_{1i}^0 + \tfrac{1}{4}(m_1 + m_j)^2 s^0 + \tfrac{1}{4}(m_2 + m_3)^2 s^0 - m_1^2 - m_2^2 - m_3^2, \qquad (18)$$

where $i = 2$, $j = 3$ for $\hat{J}_1$ and $\hat{J}_2$, $i = 3$, $j = 2$ for $\hat{J}_3$.

Such a choice of point $s^0$ allows us to replace the integral Eqs. (8) - (12) (Fig. 1) by the algebraic equations (19) - (23) respectively:

$$\alpha_1 = \lambda_1 + 2[\alpha_1 J_1 + \alpha_5 J_3]/B_2^{uu}(s_{12}^0), \qquad (19)$$

$$\alpha_2 = \lambda_2 + 2[\alpha_2 J_1^{ss} + \alpha_5 J_3^{ss}]/B_2^{ss}(s_{12}^0), \qquad (20)$$

$$\alpha_3 = \lambda_3 + 2\alpha_5 J_3^s / B_2^{us}(s_{12}^0), \qquad (21)$$

$$\alpha_4 = \lambda_4 + 2\alpha_5 J_3 / B_2^{ud}(s_{12}^0), \qquad (22)$$

$$\alpha_5 = \lambda_5 + 4[(\alpha_1 + \alpha_4)J_2 + \alpha_3 J_2^s + \alpha_2 J_2^{ss}]/[B_1(s_{12}^0)B_1(s_{34}^0)], \qquad (23)$$

These functions are similar to the functions:

$$J_1(s;m_1;m_2;m_3;m_4) = G_2^2 B_2(s_{13}^0) \int\limits_{(m_1+m_2)^2}^{(m_1+m_2)^2 \Lambda/4} \frac{ds'_{12}}{\pi} \frac{\rho_2(s'_{12})}{s'_{12}-s_{12}^0} \int\limits_{-1}^{+1} \frac{dz_1}{2} \frac{1}{1-B_2(s'_{13})}, \qquad (24)$$

$$J_2(s;m_1;m_2;m_3;m_4) = G_1^4 B_2(s_{13}^0) \times$$
$$\times \int\limits_{(m_1+m_2)^2}^{(m_1+m_2)^2 \Lambda/4} \frac{ds'_{12}}{\pi} \frac{\rho_1(s'_{12})}{s'_{12}-s_{12}^0} \int\limits_{(m_3+m_4)^2}^{(m_3+m_4)^2 \Lambda/4} \frac{ds'_{34}}{\pi} \frac{\rho_1(s'_{34})}{s'_{34}-s_{34}^0} \int\limits_{-1}^{+1} \frac{dz_3}{2} \int\limits_{-1}^{+1} \frac{dz_4}{2} \frac{1}{1-B_2(s'_{13})}, \qquad (25)$$

$$J_3(s;m_1;m_2;m_3;m_4) = G_2^2 B_1(s_{13}^0) B_1(s_{24}^0) \frac{1-B_2(s_{12}^0,\Lambda)}{1-B_2(s_{12}^0,\widetilde{\Lambda})} \frac{1}{4\pi} \times$$
$$\times \int\limits_{(m_1+m_2)^2}^{(m_1+m_2)^2 \widetilde{\Lambda}/4} \frac{ds'_{12}}{\pi} \frac{\rho_2(s'_{12})}{s'_{12}-s_{12}^0} \int\limits_{-1}^{+1} \frac{dz_1}{2} \int\limits_{-1}^{+1} dz \int\limits_{z_2^-}^{z_2^+} dz_2 \frac{1}{\sqrt{1-z^2-z_1^2-z_2^2+2zz_1z_2}} \times . \qquad (26)$$
$$\times \frac{1}{[1-B_1(s'_{13})][1-B_1(s'_{24})]}$$

$$\widetilde{\Lambda}(ik) = \begin{cases} \Lambda(ik), & \text{if } \Lambda(ik) \leq (\sqrt{s_{123}}+m_3)^2 \\ (\sqrt{s_{123}}+m_3)^2, & \text{if } \Lambda(ik) > (\sqrt{s_{123}}+m_3)^2 \end{cases} \qquad (27)$$

The other choices of point $s_0$ do not change essentially the contributions of $\alpha_1$, $\alpha_2$, $\alpha_3$, $\alpha_4$ and $\alpha_5$, therefore we omit the indices $s_0^{ik}$. Since the vertex functions depend only slightly on energy, it is possible to treat them as constants in our approximation.

The solutions of the system of equations are considered as:

$$\alpha_i(s) = F_i(s,\lambda_i)/D(s), \qquad (28)$$

where zeros of $D(s)$ determinants define the masses of bound states. $F_i(s,\lambda_i)$ determine the contributions of subamplitudes.

### III. Calculation results.

The pole of the reduced amplitudes $\alpha_1$, $\alpha_2$, $\alpha_3$, $\alpha_4$, $\alpha_5$ corresponds to the bound state and determines the mass of the cryptoexotic meson with $n=2$ and $J^{PC}=2^{++}$. The quark masses of model $m_{u,d}$ = 385 MeV, $m_s$ = 510 MeV coincide with the ordinary meson ones in our model. By means of the small changes of the masses nonstrange and strange quarks ($\overline{m}_{u,d} = m_{u,d} + \Delta$, $\overline{m}_s = m_s + \Delta$) one can effectively take into account the contribution of the confinement potential to obtaining the spectrum of exotic mesons.

The model in question has only three parameters:
(1) First group of exotic mesons (Table I): the cutoff $\Lambda = 24$, gluon coupling constant $g = 0.189$, the mass shift $\Delta = 10$ MeV.
(2) Second group of exotic mesons (Table II): the cutoff $\Lambda = 19.9$, gluon coupling constant $g = 0.218$, the mass shift $\Delta = 30$ MeV.

As usual, the masses of mesons with the spin-parity $J^{PC} = 0^{-+}$ and $J^{PC} = 1^{+-}$ coincide that they possess the similar Chew-Mandelstam functions and the vertex functions (Tables III, IV). In the first case, the parameters are determined by fixing the masses for the spin-parity $J^{PC} = 2^{++}$ M = 1565 MeV and $J^{PC} = 0^{-+}$ M = 1295 MeV. In the second case, we use the mass of the $J^{PC} = 2^{++}$ M = 1640 MeV and with the $J^{PC} = 0^{-+}$ M = 1405 MeV. The mass shifts $\Delta$ have been determined for the best description of the mass spectra of the group 1 and 2 (Tables I, II). The results are in good agreement with the experimental data [30]. We predict the masses of exotic mesons with quantum number $J^{PC} = 0^{--}, 1^{-+}$. In our approach the mass of exotic meson with the spin-parity $J^{PC} = 0^{--}$ is equal to the mass of cryptoexotic state with $J^{PC} = 0^{++}$. The interesting result of this model is the calculation of exotic meson amplitudes: the four four-quark amplitudes and the glueball amplitude. The contribution of glueball amplitudes for the first and second groups of exotic mesons are given in Tables I and II. The contributions of four-quark amplitudes are about 40 % for the first case and 50 % for the second case.

## IV Conclusion.

In a strongly bound system of light quarks such as exotic mesons, where p/m~ 1 for the light quarks, the approximation of nonrelativistic kinematics and dynamics is not justified.

The reason for the successful use of quark potential models is connected with a successful choice of the effective parameters: the mass of the quarks, the characteristics of the confinement potential, and the coupling constant $\alpha_s$. In quark models, which describe rather well the masses and static properties of light hadrons, the masses of the quarks usually have the same values for both meson and exotic mesons. However, this is achieved at the expense of some difference in the characteristics of the confinement potential. It should be borne in mind that for the fixed hadron mass the masses of dressed quarks which enter into the composition of the hadron will become smaller when the slope of the confinement potential increases or its radius decreases. Therefore conversely, one can change the masses of the quarks when going from the spectrum of ordinary mesons to the four-quark states, while keeping the characteristics of the confinement potential unchanged. For the sake of simplicity we restrict our selves to introduction of the small quark mass shift $\Delta$.

In our paper the dynamics of quark interactions is defined by the Chew-Mandelstam Functions (Tables IV). We include only three parameters: the cutoff $\Lambda$, gluon coupling constant $g$ and mass shift $\Delta$. The small quark mass shift $\Delta$ takes into account the confinement potential effectively. The relativistic four-body approach gives rise to the dynamical mixing of the four-quark amplitudes and the glueball amplitudes. We calculated the two groups of exotic mesons (Tables I, II), which are similar to the candidates for the exotica [30].

Acknowledgment.

The authors would like to thank T. Barnes, Fl. Stancu and S.-L. Zhu for useful discussions. The work was carried with the support of the Russian Ministry of Education (Grand № 2.1.1.68.26).

Fig. 1 Graphical representation of the equations for the four-quark subamplitudes $A_k$ ($k$ =1-4) and glueball subamplitude $A_5$, corresponding to the exotic mesons.

Table I. Exotic low-lying meson masses and contribution of glueball subamplitudes.
Parameters of first group mesons: cutoff $\Lambda = 24$, gluon coupling constant $g = 0.189$, quark mass shift $\Delta = 10$ MeV.
Experimental mass values of exotic mesons are given in parentheses [30].

| $J^{PC}$ | Mass, MeV | Glueball contribution % |
|---|---|---|
| $2^{++}$ | 1565 (1565) | 67.47 |
| $0^{-+} = 1^{+-}$ | 1295 (1295) | 55.27 |
| $0^{++}$ | 1422 (1370) | 60.16 |
| $1^{--}$ | 1558 (1450) | 66.93 |
| $1^{++}$ | 1500 (1420) | 63.31 |
| $1^{-+}$ | 1327 ( - ) | 56.20 |
| $0^{--}$ | 1422 ( - ) | 60.16 |

Table II. Exotic low-lying meson masses and contribution of glueball subamplitudes.
Parameters of second group mesons: cutoff $\Lambda = 19.9$, gluon coupling constant $g = 0.218$, quark mass shift $\Delta = 30$ MeV.
Experimental mass values of exotic mesons are given in parentheses [30].

| $J^{PC}$ | Mass, MeV | Glueball contribution % |
|---|---|---|
| $2^{++}$ | 1640 (1640) | 64.81 |
| $0^{-+} = 1^{+-}$ | 1405 (1405) | 52.17 |
| $0^{++}$ | 1542 (1500) | 59.15 |
| $1^{--}$ | 1632 (1570) | 64.27 |
| $1^{++}$ | 1569 (1510) | 59.95 |
| $1^{-+}$ | 1421 ( - ) | 53.13 |
| $0^{--}$ | 1542 ( - ) | 59.15 |

Table III. Vertex functions.

| $J^{PC}$ | $G_n^2$ |
|---|---|
| $1^{--}$ (n=1) | 3g |
| $2^{++}$ (n=2) | 4g/3 |
| $1^{++}$ (n=2) | 4g/3 |
| $1^{--}$ (n=2) | 4g/3 |
| $1^{+-}$ (n=2) | $8g/3 - 4g(m_i + m_k)^2/(3s_{ik})$ |
| $1^{-+}$ (n=2) | 4g/3 |
| $0^{++}$ (n=2) | $8g/3$ |
| $0^{-+}$ (n=2) | $8g/3 - 4g(m_i + m_k)^2/(3s_{ik})$ |
| $0^{--}$ (n=2) | 4g/3 |

Table IV. Coefficients of Chew-Mandelstam functions.

| $J^{PC}$ | $\alpha$ | $\beta$ | $\delta$ |
|---|---|---|---|
| $1^{--}$ (n=1) | 1/3 | 1/6-e/3 | -1/6 |
| $2^{++}$ (n=2) | 3/10 | 1/5-3e/10 | -1/5 |
| $1^{++}$ (n=2) | 1/2 | -e/2 | 0 |
| $1^{--}$ (n=2) | 1/3 | 1/6-e/3 | -1/6 |
| $1^{+-}$ (n=2) | 1/2 | -e/2 | 0 |
| $1^{-+}$ (n=2) | 2/3 | -e | 1/3 |
| $0^{++}$ (n=2) | 1/2 | -1/2 | 0 |
| $0^{-+}$ (n=2) | 1/2 | -e/2 | 0 |
| $0^{--}$ (n=2) | 1 | -1 | 0 |

$e = (m_i - m_k)^2 / (m_i + m_k)^2$

References.

1. R.L. Jaffe, Phys. Rev., D15, 281 (1977).
2. R.L. Jaffe and F.E. Low, Phys. Rev., D19, 2105 (1979).
3. J. Weinstein, N. Isgur, Phys. Rev. Lett., 48, 659 (1982), Phys. Rev., D27, 588 (1983).
4. N. Isgur and J. Paton, Phys. Rev., D31, 2910 (1985).
5. A. De Rujula, H. Georgi and S.L. Glashow, Phys. Rev., D12, 147 (1975).
6. C. Amsler and F.E. Close, Phys. Lett., B335, 425 (1995).
7. S. Godfrey and N. Isgur, Phys. Rev., D32, 189 (1985).
8. C.E. Thomas and F.E. Close, Phys. Rev., D78, 034007 (2008).
9. Y. Dong, A. Faessler, T. Gutshe and V.E. Lynbovitskij, Phys. Rev., D77, 094013 (2008).
10. Y.R. Liu, X.Liu, W.Z. Deng and S.-L. Zhu, Eur. Phys.J, C56, 63 (2008).
11. I.J.R. Aitchison, J. Phys., G 3, 121 (1977).
12. J.J. Brehm. Ann. Phys., (N.Y.) 108, 454 (1977).
13. I.J.R. Aitchison, J.J. Brehm. Phys. Rev., D17, 3072 (1978).
14. I.J.R. Aitchison, J.J. Brehm. Phys. Rev., D20, 1119 (1979).
15. J.J. Brehm. Phys. Rev., D21, 718 (1980).
16. S.M. Gerasyuta and E.F. Matskevich, Yad. Fiz., 70, 1995 (2007).
17. S.M. Gerasyuta and E.F. Matskevich, Phys. Rev., D76, 116004 (2007).
18. S.M. Gerasyuta and E.F. Matskevich, Int. J. Mod. Phys., E17, 585 (2008).
19. G. t' Hooft, Nucl. Phys., B72, 461 (1974).
20. G. Veneziano, Nucl. Phys., B117, 519 (1976).
21. E. Witten, Nucl. Phys., B160, 57 (1979).
22. O.A. Yakubovsky, Sov. J. Nucl. Phys., 5, 1312 (1967).
23. S.P. Merkuriev and L.D. Faddeev, Quantum Scattering Theory for Systems of Few Particles (Nauka, Moscow, 1985) p. 398.
24. Y. Nambu and G. Jone-Lasinio, Phys. Rev., 122,345(1961),124,246(1961).
25. T. Appelqvist and J.D. Bjorken, D4, 3726 (1971).
26. C.C. Chiang, C.B. Chiu, E.C.G. Sudarshan and X. Tata, Phys. Rev. D25, 1136 (1982).
27. V.V. Anisovich, S.M. Gerasyuta and A.V. Sarantsev. Int. J. Mod. Phys., A6, 625 (1991).
28. C.F. Chew and S. Mandelstam, Phys. Rev., 119, 467 (1960).
29. V.V. Anisovich and A.A. Anselm, Usp. Fiz. Nauk., 88, 287 (1966).
30. C. Amsler et al. (Particle Data Group), Phys. Lett., B667, 1 (2008).

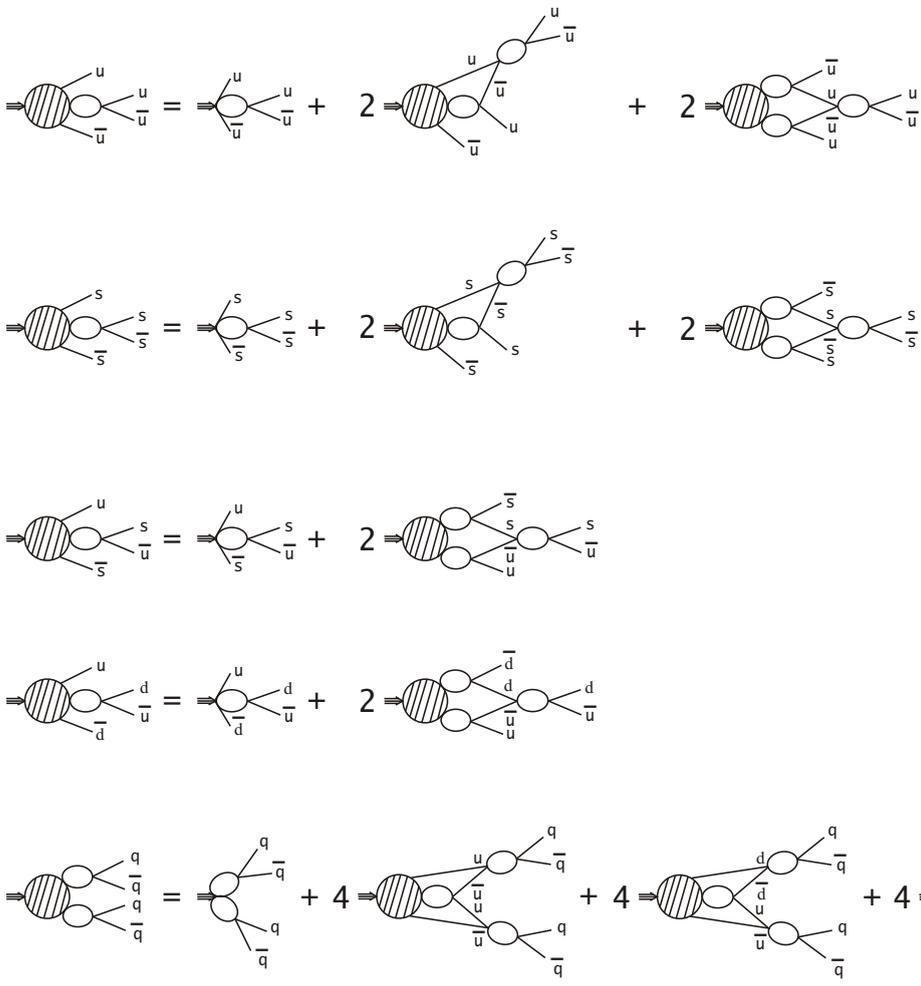

Fig. 1